\documentclass[twocolumn,aps,superscriptaddress,showpacs]{revtex4}

\usepackage{amssymb}
\usepackage{amsmath}
\usepackage{graphicx}
\usepackage[normalem]{ulem}
\usepackage[dvips]{color}

\usepackage[normalem]{ulem} 
\usepackage[dvips]{color} 
\renewcommand\sout{\bgroup \color{red} \ULdepth=-.5ex \ULset}

\begin{document}

\title{Energy dependence of pion in-medium effects on $\pi^-/\pi^+$ ratio in heavy-ion collisions}

\author{Jun Xu}
\email{xujun@sinap.ac.cn}
\affiliation{Shanghai Institute of Applied Physics, Chinese Academy
of Sciences, Shanghai 201800, China}

\author{Lie-Wen Chen}
\affiliation{Department of Physics and Astronomy and Shanghai Key
Laboratory for Particle Physics and Cosmology, Shanghai Jiao Tong
University, Shanghai 200240, China}

\author{Che Ming Ko}
\affiliation{Cyclotron Institute and Department of Physics and
Astronomy, Texas A$\&$M University, College Station, Texas 77843,
USA}

\author{Bao-An Li}
\affiliation{Department of Physics and Astronomy, Texas A$\&$M
University-Commerce, Commerce, TX 75429-3011, USA}
\affiliation{Department of Applied Physics, Xi'an Jiao Tong
University, Xi'an 710049, China}

\author{Yu-Gang Ma}
\affiliation{Shanghai Institute of
Applied Physics, Chinese Academy of Sciences, Shanghai 201800,
China}

\date{\today}

\begin{abstract}

Within the framework of a thermal model with its parameters fitted
to the results from an isospin-dependent Boltzmann-Uehling-Uhlenbeck
(IBUU) transport model, we study the pion in-medium effect on the
charged-pion ratio in heavy-ion collisions at various energies. We
find that due to the cancellation between the effects from
pion-nucleon s-wave and p-wave interactions in nuclear medium, the
$\pi^-/\pi^+$ ratio generally decreases after including the pion
in-medium effect. The effect is larger at lower collision energies
as a result of narrower pion spectral functions at lower
temperatures.
\end{abstract}


\maketitle


The density dependence of nuclear symmetry energy $E_{\rm
sym}(\rho)$ is important for understanding isospin effects in
heavy-ion reactions and properties of rare isotopes as well as many
phenomena in nuclear
astrophysics~\cite{Li98,Dan02,Lat04,Ste05,Bar05,Li08}. Extensive
studies using various probes during past many years have
significantly constrained the values of $E_{\rm sym}$ at
subsaturation densities (for a recent review, see, e.g.,
Ref.~\cite{Che12}). The behavior of $E_{\rm sum}(\rho)$ at
suprasaturation densities remains, however, uncertain in spite of
many studies based on the $\pi^-/\pi^+$
ratio~\cite{Li02,Li05,Fer06,Xiao09,Feng10,Xie13,Zha12}, $K^0/K^+$
ratio~\cite{Fer06}, n/p or t/$^3$He ratio~\cite{Li06}, and the ratio
of neutron and proton elliptic flows~\cite{Rus11} in heavy-ion
collisions as well as the properties of neutron stars~\cite{Rob12}.
In particular, contradictory conclusions were obtained on $E_{\rm
sym}(\rho)$ from the $\pi^-/\pi^+$
ratio~\cite{Li02,Li05,Fer06,Xiao09,Feng10,Xie13,Zha12}. To describe
the FOPI data from GSI, results by Xiao {\it et al.}~\cite{Xiao09}
based on the IBUU transport model together with an isospin- and
momentum-dependent interaction favors a supersoft symmetry energy,
while those of Feng {\it et al.}~\cite{Feng10} based on an improved
isospin-dependent quantum molecular dynamics model requires a stiff
symmetry energy. A more recent study by Xie {\it et
al.}~\cite{Xie13} based on an improved isospin-dependent
Boltzmann-Langevin approach also favors a supersoft symmetry energy.

In all these studies of the symmetry energy effect on
$\pi^-/\pi^+$ ratio, medium effects on the pion production threshold
in nucleon-nucleon collisions that may lead to an opposite effect
from that of the symmetry energy~\cite{Fer06}, were neglected. Also,
pions were treated as free particles in nuclear medium. It is known
that due to its coupling to the nucleon-particle-nucleon-hole and
$\Delta$-particle-nucleon-hole ($\Delta$-hole) excitations, the
dispersion relation of a pion in nuclear medium is softened and the
strength of its spectral function is enhanced at lower
energies~\cite{Bro75}. Because of the isospin dependence of
pion-nucleon interactions, the $\pi^-$ couples more strongly with
neutrons than the $\pi^+$ and thus has an even softer dispersion
relation in neutron-rich nuclear matter than $\pi^+$. On the other
hand, the pion mass in neutron-rich matter is larger for $\pi^-$
than for $\pi^+$ as a result of the pion $s$-wave interaction as
shown in Ref.~\cite{Kai01} based on the chiral perturbation
calculation. Within a thermal model for heavy-ion
collisions~\cite{Xu10}, effects from the pion-nucleon p-wave and
s-wave interactions were, however, found to largely cancel out,
resulting in only a slight reduction of the $\pi^-/\pi^+$ ratio
compared to that without the pion in-medium effects.

The pion in-medium effect studied in Ref. \cite{Xu10}
was for Au+Au collisions at only one energy of 400 AMeV, using $2\rho_0$ ($\rho_0=0.16$ fm$^{-3}$ being the
normal nuclear matter density), $T=43.6$ MeV, and
$\delta_{\rm like}=0.135$ for the density, temperature, and isospin
asymmetry of produced matter. Since different densities and
temperatures are reached in heavy-ion collisions at different beam
energies, the $\pi^-/\pi^+$ ratio is expected to be modified
differently by pion in-medium effects. It is therefore worth to
investigate the beam energy dependence of pion in-medium effects to
find the most suitable collision energies where the $\pi^-/\pi^+$
ratio is more sensitive to the density dependence of nuclear
symmetry energy than the pion in-medium effects. In this paper, we carry out such a study within a similar framework as
in Ref.~\cite{Xu10} but using thermal model parameters that are more
accurately determined from the IBUU transport model calculations.



In high-energy heavy-ion collisions, pions and $\Delta$ resonances
are mostly produced from the initial high-density ($\rho>\rho_0$)
phase. Assuming that this high-density matter is in thermal
equilibrium with baryon density $\rho_B=\rho_n + \rho_p +
\rho_\Delta^{-} + \rho_\Delta^{0} + \rho_\Delta^{+} +
\rho_\Delta^{++}$, isospin asymmetry $\delta_{\rm
like}=(\rho_n-\rho_p+\rho_{\Delta^-}-\rho_{\Delta^{++}}+\rho_{\Delta^0}/3-\rho_{\Delta^+}/3)/\rho_B$,
and temperature $T$. The densities of nucleons, $\Delta$ resonances,
and pions with isospin states $m_\tau$, $m_T$, and $m_t$ can then be
expressed, respectively, as \small
\begin{eqnarray}\label{density}
\rho_N^{m_\tau} &=& 2 \int \frac{d^3{\bf p}}{(2\pi)^3} \frac{1}{e^{(m_N^{}
+ p^2/2m_N^{} +U_N^{m_\tau}-\mu_B^{}-2m_\tau\mu_Q^{})/T}+1}, \notag \\
\rho_\Delta^{m_T} &=& 4 \int \frac{d^3{\bf p}}{(2\pi)^3}
\frac{P_\Delta^{m_T}(M) dM}{z_\Delta^{-1}e^{[M + p^2/2M
+U_\Delta^{m_T}-\mu_B^{}-(m_T+\frac{1}{2})\mu_Q^{}]/T}+1}, \notag\\
\rho_\pi^{m_t} &=& \int \frac{d^3{\bf p}}{(2\pi)^3} \frac{
S_\pi^{m_t }(\omega,p)d \omega^2}{z_\pi^{-1}e^{(\omega -
m_t\mu_Q^{})/T}-1}.
\end{eqnarray}
\normalsize In the above, $U_N^{m_\tau}$ is the momentum-independent
nucleon mean-field potential with an isoscalar as well as an
isovector part. The former is fitted to the binding energy $E_0=-16$
MeV and incompressibility $K_0=230$ MeV of normal nuclear matter at
saturation density $\rho_0$. The latter is taken to be from either
the supersoft symmetry energy that vanishes at about $3\rho_0$ and is
denoted by $x=1$ or a stiff one that increases almost linearly
with density and is denoted by $x=0$ in Ref.~\cite{Che05a}. These
symmetry energies correspond to current uncertainties in $E_{\rm
sym}(\rho)$ at suprasaturation densities. $U_\Delta^{m_T}=
\sum_{m_\tau,m_t^{}}\left|\left\langle {\textstyle\frac{3}{2}}\,
m_T^{} | 1\, m_t\, {\textstyle\frac{1}{2}} \, m_\tau \right\rangle
\right|^2 U_N^{m_\tau}$ is the $\Delta$ mean-field potential with
$\langle {\textstyle\frac{3}{2}} \, m_T^{} | 1\,
m_t\,{\textstyle\frac{1}{2}} \, m_\tau \rangle$ being the
Clebsch-Gordan coefficient from the isospin coupling of the $\Delta$
resonance with nucleon and pion.  $\mu_B$ and $\mu_Q$ are the baryon
and charge chemical potentials, respectively. $P_\Delta^{m_T}(M)$
and $S_\pi^{m_t }(\omega,p)$ are the mass distribution of $\Delta$
resonance and the pion spectral function including effects due to
both pion-nucleon s-wave and p-wave interactions, and they are
determined from a self-consistent calculation as described in
details in Ref.~\cite{Xu10}.

\vspace{-0.2cm}
\begin{figure}[h]
\centerline{\includegraphics[scale=0.8]{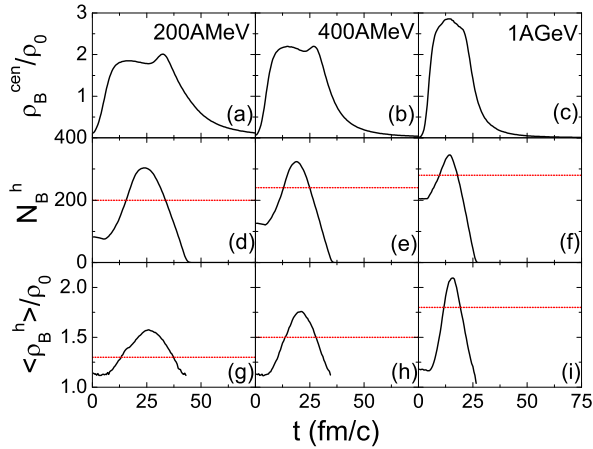}} \caption{(Color
online) Time evolution of central baryon density (top panels),
baryon multiplicity (middle panels), and average baryon density
(bottom panels) in the high-density $(\rho>\rho_0)$ phase of central
Au+Au collisions at different beam energies. Horizontal lines indicate the
average values used in the thermal model.} \label{ibuu}
\end{figure}

The $z_\pi$ and $z_\Delta$ in Eq.~(\ref{density}) are the fugacity
parameters of pions and $\Delta$ resonances, respectively, and their
values at a given temperature depend on the pion and $\Delta$
resonance numbers relative to their equilibrium values. Because of
the short chemical equilibration times estimated from the pion and
$\Delta$ resonance production rates in the absence of in-medium
effects, we assume that both $z_\pi$ and $z_\Delta$ are 1 in this
case. To determine their values when in-medium effects are included,
we first determine the baryon density, isospin asymmetry, and
temperature of the nuclear medium from the IBUU transport model
calculations with the same nucleon and $\Delta$ resonance mean-field
potentials as used in the thermal model but without the pion
in-medium effects. Figure~\ref{ibuu} displays the time evolution of
the central baryon density, the baryon multiplicity, and the average
baryon density in the high-density phase of central Au+Au collisions
at three beam energies of 200 AMeV, 400 AMeV, and 1 AGeV. Higher
central densities are reached at higher collision energies but last
a shorter duration {than at lower collision energies. Also, the
baryon number in the high-density region changes with time, and the
average values are about 200 for 200 AMeV, 240 for 400 AMeV, and 280
for 1 AGeV as given in Table~\ref{tab}. They are, however, similar
when different $E_{\rm sym}(\rho)$ are used in the IBUU model.

\vspace{-0.5cm}
\begin{table}[h]
\caption{{\protect\small Baryon number $N_B^h$, pion-like particle
multiplicity $N_\pi$, their ratio $N_\pi/N_B^h$, average baryon
density $\langle \rho^h_B \rangle$, $\pi^-/\pi^+$ ratio from the
IBUU model with free pions and symmetry energy $x=1$ ($x=0$) for
central Au+Au collisions at different beam energies; and temperature
$T$, isospin asymmetry $\delta_{\rm like}$, baryon and charge
chemical potentials $\mu_B$ and $\mu_Q$, fugacity parameters
$z_\pi^{}$ and $z_\Delta^{}$ for pions and $\Delta$ resonances, and
$\pi^-/\pi^+$ ratio with pion in-medium effects in the corresponding
thermal model.}} \label{tab}
\begin{tabular}{ccccccc}
\hline\hline
$E_{\rm beam}$  & 200 AMeV \quad & 400 AMeV \quad & 1 AGeV \\
\hline
$$ $N_B^h$ & 200 & 240 & 280 \\
$$ $N_\pi$ & 0.52 (0.47) & 9.59 (9.25) & 70.8 (70.2) \\
$$ $N_\pi/N_B^h$ ($\times10^{-3}$) & 2.60 (2.36) & 40.0 (38.5) & 253 (251)\\
$$ $\langle \rho^h_B \rangle/\rho_0$ & 1.3 & 1.5 & 1.8 \\
$$ free $\pi^-/\pi^+$ & 4.34 (4.05) & 2.74 (2.63) & 1.79 (1.76)\\
\hline
$$ $T$ (MeV) & 32.6 (32.2) & 53.3 (52.9) & 90.8 (90.5)\\
$$ $\delta_{\rm like}$ & 0.152 (0.128) & 0.163 (0.136) & 0.211 (0.185)\\
$$ $\mu_B$ (MeV) & 919.9 (920.6) & 894.3 (895.8) & 814.9 (817.0)\\
$$ $\mu_Q$ (MeV) & -23.6 (-22.5) & -24.8 (-24.5) & -23.1 (-23.8)\\
$$ $z_\pi^{}$ & 0.115 (0.116) & 0.190 (0.189) & 0.392 (0.391)\\
$$ $z_\Delta^{}$ & 0.422 (0.417) & 0.667 (0.634) & 0.813 (0.806) \\
$$ in-med. $\pi^-/\pi^+$ & 4.15 (3.85) & 2.70 (2.59) & 1.80 (1.77) \\
\hline\hline
\end{tabular}%
\end{table}

The multiplicities of pion-like particles at different collision
energies are also given in Table~\ref{tab}, and they are slightly
larger for $x=1$ than for $x=0$. For the temperature of the
high-density phase, it is determined from fitting the ratio of the
pion-like particle number to the baryon number using the thermal
model. Since the average baryon density of the high-density phase
changes with time as shown in the bottom panels of Fig.~\ref{ibuu},
we use the mean values of $1.3\rho_0$ for 200 AMeV, $1.5\rho_0$ for
400 AMeV, and $1.8\rho_0$ for 1 AGeV for the thermal model
calculations.  As expected, the resulting temperature and
multiplicity of pion-like particles in the high-density phase
increase with increasing collision energy as shown in
Table~\ref{tab}. For the isospin asymmetry $\delta_{\rm like}$ used
in the thermal model, it is determined by fitting the final
$\pi^-/\pi^+$ ratio from the IBUU calculations shown in
Table~\ref{tab}. It is seen that $\delta_{\rm like}$ is larger for
$x=1$, consistent with the argument that a softer symmetry energy
leads to a more neutron-rich high-density phase and thus a larger
$\pi^-/\pi^+$ ratio as $\pi^{-}(\pi^+)$ is mostly produced from
neutron-neutron (proton-proton) scatterings.

Further shown in Table~\ref{tab} are the baryon and charge chemical
potentials $\mu_B$ and $\mu_Q$ determined from the chemical
equilibrium conditions. Since the multiplicity of pion-like
particles is much smaller than the total nucleon number in heavy-ion
collisions at energies considered here, including pion in-medium
effects has negligible effects on the extracted values for $\rho_B$,
$\delta_{\rm like}$, and $T$ as the dynamics of heavy-ion collisions
is hardly modified. Indeed, the multiplicities of final pions at
different collision energies are reasonably reproduced by transport
model calculations with free pions~\cite{Xiao09,Feng10,Xie13}, and
including pion in-medium effects does not affect much the total pion
yield~\cite{Xio93}. Therefore, same values of $\mu_B$ and $\mu_Q$
are obtained with and without the pion in-medium effects. However,
due to the softening of pion spectral functions and the broadening
of $\Delta$ mass distributions from pion in-medium effects, the
multiplicities of pion-like particles would increase if they are
assumed to be in chemical equilibrium with nucleons. As in
Ref.~\cite{Xu10}, to reproduce the pion and $\Delta$ resonance
multiplicities from the IBUU model with free pions, their fugacity
parameters used in the thermal model become much smaller than one if
pion in-medium effects are included as shown in Table~\ref{tab}.
Their values increase, however, with increasing collision energy
due to a shorter chemical equilibration time.


\begin{figure}[h]
\centerline{\includegraphics[scale=0.8]{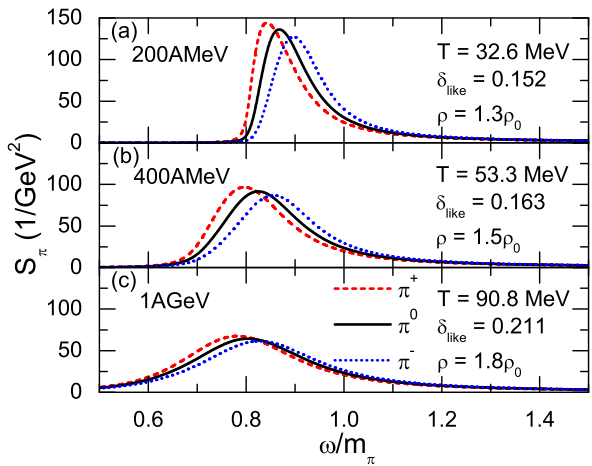}} \caption{(Color
online) Pion spectral functions
at momentum $p=m_\pi$
with the supersoft symmetry energy $x=1$.} \label{sp}
\end{figure}

In Fig.~\ref{sp}, we show the pion spectral functions at a typical
pion momentum $p=m_\pi$, obtained with the supersoft symmetry energy $x=1$,
in the presence of pion-nucleon s-wave and
p-wave interactions at different temperatures corresponding to
different collision energies of 200 AMeV, 400 AMeV, and 1 AGeV. It
is seen that the $\pi^+$ spectral function peaks at lower energies
compared to that of $\pi^-$, resulting in an increase of the $\pi^+$
multiplicity and a decrease of that of $\pi^-$. Also, the widths of
pion spectral functions become broader with increasing temperature,
leading to a larger isospin-dependent pion in-medium effects at
lower collision energies.

\begin{figure}[h]
\centerline{\includegraphics[scale=0.8]{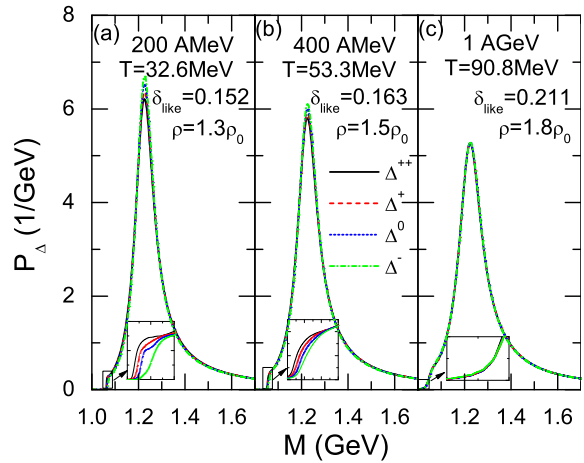}} \caption{(Color
online) Same as Fig.~\ref{sp} for the $\Delta$ resonance mass
distributions.} \label{pd}
\end{figure}

The temperature dependence of the pion spectral functions is related
to that of $\Delta$ mass distributions, as displayed in
Fig.~\ref{pd} for the supersoft symmetry energy $x=1$ as well. Similar to
the pion spectral functions, the $\Delta$
mass distribution is broader at higher temperatures. At lower
temperatures or collision energies, there is a larger probability
for $\Delta^{++}$ to have lower masses than for $\Delta^{-}$ as
shown in the insets. As a result, the production of positive
pion-like particles is more enhanced than that of negative ones.
This isospin-dependent in-medium effect is, however, rather small at
higher temperatures. Since pions in the IBUU transport model are
produced from $\Delta$ resonance decay, the mass distribution of
$\Delta$ resonances near the nucleon and pion threshold is
important for determining the pion yield in heavy-ion collisions
near and below the threshold energy.

The final $\pi^-/\pi^+$ ratios after including both pion-nucleon
s-wave and p-wave interactions at different collision energies with
different $E_{\rm sym}(\rho)$ are given in Table~\ref{tab} and
compared to their values using free pions in the left panel of
Fig.~\ref{ratio}. It is seen that the pion in-medium effects
generally reduce the $\pi^-/\pi^+$ ratio, leading to the need of an
even softer symmetry energy for reproducing their ratio using free
pions. In addition, although the $\pi^-/\pi^+$ ratio at lower
collision energies is larger and more sensitive to the symmetry
energy in the absence of medium modifications of the production
threshold, the pion in-medium effects are also larger, especially at
energies below the pion production threshold. This is consistent
with the stronger isospin-dependent pion in-medium effects at lower
temperatures shown in Figs.~\ref{sp} and \ref{pd}.

\begin{figure}[h]
\centerline{\includegraphics[scale=0.8]{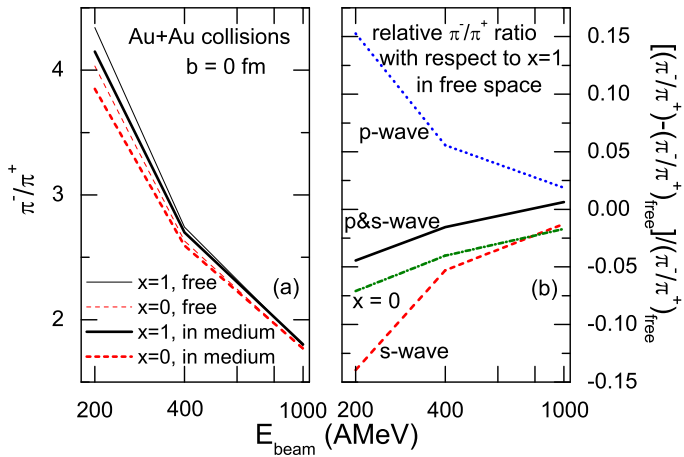}} \caption{(Color
online) Collision energy dependence of (a) $\pi^-/\pi^+$ ratios with
and without pion in-medium effects and (b) relative $\pi^-/\pi^+$
ratios from pion-nucleon s-wave interaction, p-wave interaction,
both p- and s-wave interactions, and $x=0$ with respect to that from
$x=1$ in free space from thermal model for central Au+Au collisions
at different energies. } \label{ratio}
\end{figure}

To understand the relative effect due to the pion-nucleon s-wave and
p-wave interactions,  we compare in the right panel of
Fig.~\ref{ratio} the $\pi^-/\pi^+$ ratios due to the symmetry energy
effect with those due to only the s-wave interaction, only the
p-wave interaction, and both interactions. With only the s-wave
interaction, the $\pi^-/\pi^+$ ratio is significantly reduced. At
the collision energy of 200 AMeV, this reduction is about $14\%$ and
is larger than the symmetry energy effect of about $7\%$. Including
only the p-wave interaction, the $\pi^-/\pi^+$ ratio is, on the
other hand, increased appreciably, reaching about $15\%$ at the
collision energy of 200 AMeV. After including both s-wave and p-wave
interactions, the effect of the s-wave interaction turns out to
dominate over that of the p-wave interaction, leading to a $4\%$
reduction in the $\pi^-/\pi^+$ ratio. This value is still comparable
to that due to the symmetry energy effect. At higher collision
energies, the pion in-medium effect on the $\pi^-/\pi^+$ ratio
becomes, however, very small.


To summarize, we have found via a thermal model that the pion
in-medium effects reduce the $\pi^-/\pi^+$ ratio in high-energy
heavy-ion collisions compared to that using free pions in spite of
the cancellation between the pion-nucleon s-wave interaction, which
reduces the $\pi^-/\pi^+$ ratio, and the p-wave interaction, which
increases the $\pi^-/\pi^+$ ratio. Although at lower energies the
charged-pion ratio is more sensitive to the symmetry energy, the
pion in-medium effect is also larger, especially at collision
energies below the pion production threshold. Our results thus
indicate that to understand quantitatively the symmetry energy
effect on pion production in heavy-ion collisions, it is important
to include the isospin-dependent pion in-medium effects, although
this is highly nontrivial in the transport model.


This work was supported by the "100-talent plan" of Shanghai
Institute of Applied Physics under grant Y290061011 from the Chinese
Academy of Sciences, the US National Science Foundation under Grant
No. PHY-1068572 and PHY-1068022, the Welch Foundation under Grant
No. A-1358, the National Aeronautics and Space Administration under
grant NNX11AC41G issued through the Science Mission Directorate, the
CUSTIPEN (China-U.S. Theory Institute for Physics with Exotic
Nuclei) under DOE grant number DE-FG02-13ER42025, the NNSF of China
(11135011, 11275125, 11035009, and 11220101005), the Shanghai
Rising-Star Program (11QH1401100), Shanghai "Shu Guang" Project, the
Eastern Scholar Program, and the STC of Shanghai Municipality
(11DZ2260700).

\end{document}